\documentclass[conference]{IEEEtran}
\IEEEoverridecommandlockouts
\usepackage{url}
\usepackage{graphicx}
\usepackage{booktabs}
\usepackage{cite}
\usepackage{times}
\usepackage[margin=1in]{geometry}
\usepackage{tikz}
\usetikzlibrary{shapes.geometric, arrows, positioning, fit}

\tikzstyle{startstop} = [rectangle, rounded corners, minimum width=2.5cm, minimum height=0.8cm, text centered, draw=black, fill=red!30]
\tikzstyle{process} = [rectangle, minimum width=2.5cm, minimum height=0.8cm, text centered, draw=black, fill=orange!30]
\tikzstyle{decision} = [diamond, minimum width=2cm, minimum height=0.8cm, text centered, draw=black, fill=green!30]
\tikzstyle{arrow} = [thick,->,>=stealth]

\begin{document}

\title{AI-Driven Real-Time Kick Classification in Olympic Taekwondo Using Sensor Fusion}

\author{\IEEEauthorblockN{Jamsheed Mistri}
\IEEEauthorblockA{University of California, Berkeley\\
jmistri@berkeley.edu}
}

\maketitle

\begin{abstract}
Olympic Taekwondo has faced challenges in spectator engagement due to static, defensive gameplay and contentious scoring. Current Protector and Scoring Systems (PSS) rely on impact sensors and simplistic logic, encouraging safe strategies that diminish the sport’s dynamism. This paper proposes an AI-powered scoring system that integrates existing PSS sensors with additional accelerometers, gyroscopes, magnetic/RFID, and impact force sensors in a sensor fusion framework. The system classifies kicks in real-time to identify technique type, contact location, impact force, and even the part of the foot used. A machine learning pipeline employing sensor fusion and Support Vector Machines (SVMs) is detailed, enabling automatic kick technique recognition for scoring. We present a novel kick scoring rubric that awards points based on specific kick techniques (e.g., turning and spinning kicks) to incentivize dynamic attacks. Drawing on a 2024 study achieving 96--98\% accuracy, we validate the feasibility of real-time kick classification and further propose enhancements to this methodology, such as ensemble SVM classifiers and expanded datasets, to achieve the high-stakes accuracy required by the sport. We analyze how the proposed system can improve scoring fairness, reduce rule exploitation and illegitimate tactics, encourage more dynamic techniques, and enhance spectator understanding and excitement. The paper includes system design illustrations, a kick scoring table from an AI-augmented rule set, and discusses anticipated impacts on Olympic Taekwondo.
\end{abstract}


\section{Introduction}
Olympic Taekwondo has in recent years been criticized for declining spectator appeal and static gameplay. Matches often feature cautious tactics where competitors refrain from using flashy techniques, leading to what some describe as ``foot fencing,'' a defensive style dominated by light front-leg kicks. This problem has been attributed partly to limitations of the current electronic Protector and Scoring System (PSS) and the scoring rules in place. After Taekwondo’s inclusion as an Olympic medal sport, the International Olympic Committee (IOC) pressured World Taekwondo (WT) to improve the sport’s objectivity and entertainment value, as early matches were deemed too boring for audiences \cite{Moenig2015}. In the early 2000s, WT introduced electronic scoring to replace subjective judging, hoping to eliminate bias and encourage more action. While objectivity improved, the unintended consequence was a shift toward a constrained fighting style focused on safely triggering electronic sensors rather than delivering dynamic techniques \cite{Moenig2015}. Athletes learned to exploit the PSS by using repetitive, low-risk techniques (often quick tap kicks with the front foot) that reliably score points under the system’s thresholds. This ``safety-first'' approach, combined with complex scoring rules, has made it difficult for casual spectators to appreciate the sport’s full excitement. As one observer noted, Taekwondo under current rules ``has never been truly fit for television'' because new viewers struggle to follow the scoring and see little spectacular action \cite{Sexton2022}. 

The static gameplay not only reduces fan engagement but also conflicts with Taekwondo’s martial heritage of explosive, high-flying kicks. Recognizing this, officials and technologists have sought solutions. For example, next-generation PSS designs explicitly aim to \emph{“reduce static taekwondo, where primarily front kicks are used, and reward athletes who practice more dynamic taekwondo”} \cite{Daedo2023}. This paper addresses the issue by proposing an AI-enhanced scoring system that leverages sensor fusion and machine learning to recognize a wide array of kick techniques in real time. By intelligently identifying and scoring diverse kicks, the system can incentivize fighters to employ more dynamic strategies, thereby making matches more exciting and fair. We hypothesize that augmenting the PSS with additional sensors and AI analytics will both restore the spectacle of Taekwondo’s flying kicks and ensure that points are awarded more equitably.

The remainder of this paper is organized as follows. In Section II, we review the current PSS technology and its limitations that contribute to static play. Section III introduces the proposed sensor fusion system design, describing how various sensors (inertial, magnetic, and force) are deployed to capture detailed kick information. Section IV outlines the machine learning pipeline, including data processing and an SVM-based classifier for kick technique recognition. Section V presents a new kick scoring rubric that allocates point values by technique difficulty. Section VI describes experimental validation and draws on prior work, including a 2024 study by Liu \emph{et al.}, to demonstrate the feasibility of accurate kick classification. Section VII discusses anticipated effects of the new system on fairness, gameplay, and spectator engagement. Finally, Section VIII concludes the paper. An acknowledgments section is included to credit contributions.

\section{Current PSS and Limitations}
The Protector and Scoring System (PSS) currently used in Olympic Taekwondo consists of electronic body protectors and headgear worn by athletes, along with sensor-equipped socks/foot protectors. The two main officially approved systems, by Daedo and KPNP, use a combination of impact sensors and wireless transmission to automatically register scoring techniques. A typical PSS setup for scoring to the trunk (body) includes an array of pressure sensors or piezoelectric pads in the chest protector that measure the force of impact. For head kicks, a common solution is a magnetic sensor: the athlete’s socks contain magnets or RFID tags in the foot instep and heel, and the headgear has corresponding magnetic field sensors. When a foot comes sufficiently close to the headgear with enough force, the system registers a head kick. Both trunk and head sensors have preset calibration thresholds; a hit must exceed a force level (or sensor signal level) to count as a point \cite{Choi2021}. This threshold system was introduced to enforce objectivity and consistency in scoring, replacing the old subjective judging by corner judges.

The current scoring rules under PSS (as of the 2020s) assign points based on the location and technique of the hit. According to the standard WT competition rules \cite{WT2024}, a valid punch to the body is 1 point, a basic kick to the body is 2 points, and a basic kick to the head is 3 points. More challenging techniques earn bonus points: a turning (spinning) kick that lands on the body protector earns 4 points (2 base + 2 bonus), and a turning kick to the head earns 5 points (3 base + 2 bonus). These values are summarized as the “Current Scoring System” in Table~\ref{tab:currentscoring} for reference. The rationale is to reward higher-risk, more difficult maneuvers (spinning kicks) with additional points. In practice, however, the PSS cannot \emph{itself} distinguish whether a body kick was executed with a spin or not; instead, a human referee presses a button to manually add technical bonus points when they observe a turning kick. This reliance on human judgment for recognizing technique can introduce subjectivity and inconsistency, partially undermining the objectivity that electronic scoring aims for.

\begin{table}[t]
\centering
\caption{Current Standard Scoring for Valid Techniques (World Taekwondo rules)}
\label{tab:currentscoring}
\begin{tabular}{l c}
\toprule
\textbf{Technique} & \textbf{Points Awarded} \\
\midrule
Punch (to trunk) & 1 \\
Basic Body Kick & 2 \\
Basic Head Kick & 3 \\
Turning (Spinning) Body Kick & 4 \\
Turning (Spinning) Head Kick & 5 \\
\bottomrule
\end{tabular}
\end{table}

Despite the technological advancements of PSS, several limitations contribute to static gameplay and perceived unfairness:
\begin{itemize}
    \item \textbf{Threshold and Calibration Issues:} The force thresholds are fixed or set per weight category and can be somewhat arbitrary \cite{Choi2021}. If set too high, lighter or faster kicks that are skillfully delivered might not register, frustrating athletes. If set too low, accidental or grazing touches score points. Studies have found that current threshold settings may not accurately reflect differences in athlete size or the true scoring impact needed \cite{Choi2021}. Moreover, athletes and coaches sometimes doubt the consistency of these thresholds, leading to conservative tactics to ensure clearly registering hits.
    \item \textbf{Limited Technique Recognition:} The PSS sensors only detect that a valid impact occurred on a scoring area (trunk or head) and exceed a force threshold; they do not identify what kick was used. This means that an easy, safe kick (e.g. a light push kick) counts the same as a more spectacular technique (e.g. a jumping spinning kick) if both manage to trigger the sensors. The only differentiation is the binary spinning bonus added by referees when they notice a turn. The system does not detect finer distinctions like whether a kick was a side kick vs. a roundhouse, or whether the heel or the ball of the foot made contact. Because the electronics treat all triggering hits uniformly, athletes optimize for the simplest kick that can score. This has led to the prevalence of the front-leg side kick or ``cut kick'' used repetitively to tap the opponent’s body protector just hard enough to score, instead of riskier turning or jumping kicks.
    \item \textbf{Defensive Exploits and Static Exchanges:} With scoring heavily reliant on sensor activation, athletes often keep one leg up (to quickly poke the opponent’s protector and also to block incoming kicks) for extended periods. This static posture, colloquially called \emph{foot fencing}, yields slow-paced bouts with minimal full-body movement. The PSS encourages this because multiple quick taps with the front foot can accumulate points while minimizing exposure to counterattacks. It has been observed that many Olympians now avoid extended combinations or big motions; they score by carefully placing medium-force kicks to score just above the threshold, then retreat or clinch. Such patterns make matches appear less dynamic and entertaining than in pre-electronic scoring days.
    \item \textbf{Potential Missed Calls and Gaming the System:} Since PSS only registers certain contact conditions, some legitimate techniques might not score due to sensor limitations (for example, a perfectly timed hook kick that lands with the sole of the foot might not trigger the sensor if not enough normal force is applied, even if it is impressively skillful). Conversely, there have been cases of false positives or controversial scoring (e.g., a strike barely grazing the headgear magnet registering as a full head kick). Athletes have adapted by learning how to ``game'' the PSS, for instance by sliding their foot on the protector to create sensor pressure or exploiting clinch situations to deliver short scoring taps that the opponent cannot avoid. These behaviors, while within the rules, are viewed as undermining the spirit of Taekwondo.
\end{itemize}

In summary, the current PSS has improved objectivity but is inherently limited in the richness of data it collects about each hit. It reduces the complex martial art of Taekwondo into a few sensor signals, missing important context like the technique performed or whether the hit was clean or glancing. This gap between what is measured and what spectators value (technique difficulty, dynamic action) leads to a disconnect. There is a clear need for a smarter scoring system that not only detects contact and force, but also understands the \emph{quality and type of technique}, so that scoring can align better with the skill and excitement demonstrated by the athletes. The next section proposes a comprehensive sensor fusion approach to address these limitations.

\section{Sensor Fusion and System Design}
\begin{figure}[htb]
\centering
    \includegraphics[width=1\linewidth]{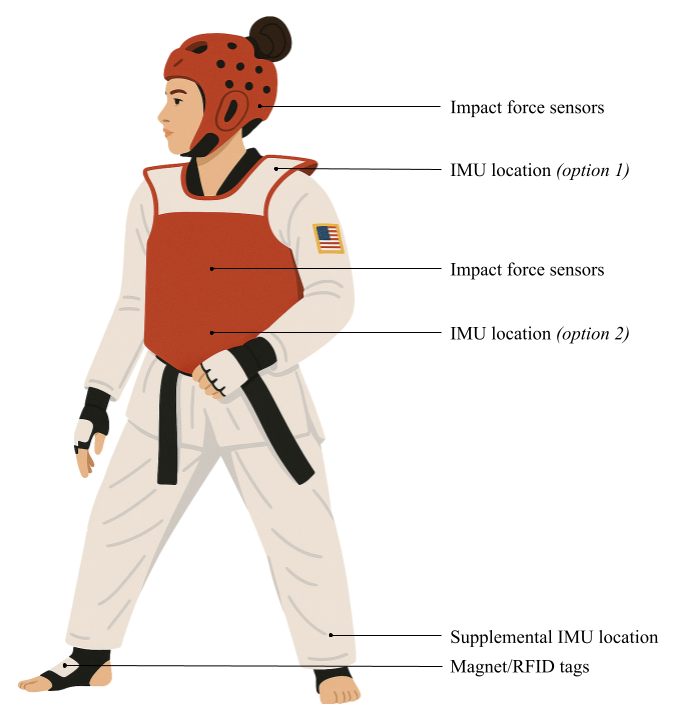}
\caption{Illustration of proposed sensor placement on an athlete's gear. Accelerometers and gyroscopes (IMUs) are attached on the abdomen and ankles/feet. The existing PSS impact sensors in the trunk protector and headgear measure force of contact. Magnet/RFID tags on the feet and corresponding sensors in the protectors detect contact location and foot part. This multi-sensor setup provides rich data for classifying kicks. (Note: actual implementation may use wireless IMU modules and integrate with current protector sensors.)}
\label{fig:sensorplacement}
\end{figure}

To capture the full context of a Taekwondo kick, we propose augmenting the traditional PSS with a network of additional sensors on the athlete’s gear. These include inertial measurement units (IMUs) such as accelerometers and gyroscopes on the athlete’s body and legs, as well as enhanced contact sensors like distributed magnet/RFID tags and pressure sensors on both the striking foot and the target zones. By fusing data from multiple sensor modalities, the system can infer not just that a hit occurred, but \emph{how} it occurred. Figure~\ref{fig:sensorplacement} illustrates the envisioned placement of sensors on an athlete:
\begin{itemize}
    \item \textbf{Impact Force Sensors:} The existing pressure sensors in the trunk protector and headgear remain in use to measure the force of impact when a kick lands. These sensors provide a quantitative measure of strike strength. In our enhanced system, we assume these sensors can output a force value or at least a calibrated score proportional to impact. This will be used both to determine if a hit meets minimum force criteria and to scale scoring for especially powerful strikes (if desired by future rules).
    \item \textbf{Magnetic/RFID Proximity Sensors:} We include the magnetic sensor mechanism from current systems: the athlete’s socks contain several small magnets or RFID tags distributed across the foot (toes, instep, heel) \cite{Daedo2023}. Corresponding magnetic field sensors or RFID receivers are embedded in the opponent’s protectors (chest and helmet). When the foot comes into close proximity (within a few centimeters) of the protector, the system registers a contact event and can identify which region was hit (e.g., anterior abdominal wall vs. obliques) based on which sensor was triggered. This provides the contact location. Additionally, having multiple magnet positions on the foot can help determine which part of the foot made the contact: for instance, if a magnet near the heel triggers the sensor, it implies a heel strike, whereas a magnet near the toes suggests a kick with the ball of the foot or instep. Such information can be useful in distinguishing certain techniques (a hook kick uses the heel, a front kick uses the ball of foot, etc.).
    \item \textbf{Accelerometers and Gyroscopes:} IMUs are attached to key points on the athlete’s body. At minimum, we propose an accelerometer/gyro unit on the waist (hip area) or upper back (trapezius shoulder area), with optional additional units on each foot (or ankle). The abdominal IMU captures the core motion of the torso and hips, which is highly indicative of the type of kick being executed (since different kicks involve distinct hip rotations and linear motions) \cite{Daedo2023}. The foot IMUs capture the fine-grained motion of each kicking leg, including speed, trajectory, and orientation of the foot. For example, a roundhouse kick (dolyo chagi) will have a very different angular velocity profile and foot trajectory than a side kick (yeop chagi). The combination of acceleration and rotational velocity data from these sensors can be used to classify the kick technique through machine learning, as will be detailed in Section~IV. Notably, prior research has demonstrated that even a single waist-mounted accelerometer can achieve high accuracy in distinguishing basic kicks \cite{Liu2024}. Our system leverages multiple IMUs for potentially finer discrimination (especially for complex kicks or to tell apart similar kicks like a rear leg versus front leg roundhouse).
    \item \textbf{Microcontroller and Wireless Unit:} All sensor data is collected in real-time by a wearable microcontroller system (which could be integrated into the athlete’s belt or protective gear). This unit handles synchronization of multi-sensor data streams (ensuring that accelerometer, gyro, and impact sensor data can be correlated in time), performs some local preprocessing, and wirelessly transmits the sensor fusion data to the scoring computer. Modern wireless protocols (like Bluetooth Low Energy or a dedicated RF channel as used in existing PSS) can be employed for transmission with minimal latency.
    \item \textbf{Scoring Computer with AI Module:} The central component of the system is a computer (edge processor) connected to the competition scoring software. This computer receives the sensor data, runs the real-time machine learning algorithms to classify the kick and determine the appropriate score, and then outputs scoring decisions to the display system for judges and spectators. The AI module is the part that interprets sensor patterns to recognize the kick type and quality.
\end{itemize}

\begin{figure*}[htb]
    \centering
  \begin{tikzpicture}[node distance=0.8cm and 1.2cm,
                      box/.style={rectangle,draw,fill=blue!10,
                                  text width=2.2cm,minimum height=.8cm,
                                  font=\footnotesize,text centered},
    sensor/.style={rectangle, draw, fill=green!10, text width=1.8cm, text centered, minimum height=0.6cm, font=\tiny},
    arrow/.style={->, thick},
    data/.style={rectangle, draw, dashed, fill=yellow!10, text width=2cm, text centered, minimum height=0.6cm, font=\tiny}
]

\node[sensor] (impact) {Impact Sensors (Trunk/Head)};
\node[sensor, right=of impact] (magnetic) {Magnetic/RFID Sensors};
\node[sensor, right=of magnetic] (imu) {IMU Sensors (Accel/Gyro)};

\node[box, below=of magnetic] (transmitter) {Wireless Transmitter Module};

\node[box, below=1.2cm of transmitter] (ai) {AI Processing Unit (Edge Computer)};

\node[data, left=of ai] (classify) {Real-time Kick Classification};
\node[data, right=of ai] (verify) {Impact Verification};

\node[box, below=1.2cm of ai] (scoring) {Scoring System Software};

\node[box, left=of scoring] (display) {Score Display};
\node[box, right=of scoring] (referee) {Referee Interface};

\draw[arrow] (impact) -- (transmitter);
\draw[arrow] (magnetic) -- (transmitter);
\draw[arrow] (imu) -- (transmitter);

\draw[arrow] (transmitter) -- node[right, font=\tiny] {Sensor Fusion Data} (ai);

\draw[arrow] (ai) -- (classify);
\draw[arrow] (ai) -- (verify);
\draw[arrow] (classify) -- (ai);
\draw[arrow] (verify) -- (ai);

\draw[arrow] (ai) -- node[right, font=\tiny] {Kick Type + Points} (scoring);

\draw[arrow] (scoring) -- (display);
\draw[arrow] (scoring) -- (referee);

\end{tikzpicture}

\caption{System integration schematic for the AI-enhanced PSS. The athlete's wearable sensors (impact, magnetic, IMUs) feed into a local transmitter module, which sends fused data to the AI processing unit. The AI unit (edge computer) runs real-time classification to identify the kick type and verifies impact. It then communicates with the scoring system software to automatically assign points per the new rubric. The entire cycle from sensor trigger to score display is designed to occur within fractions of a second, minimizing any delay in scoring feedback.}
\label{fig:integration}
\end{figure*}
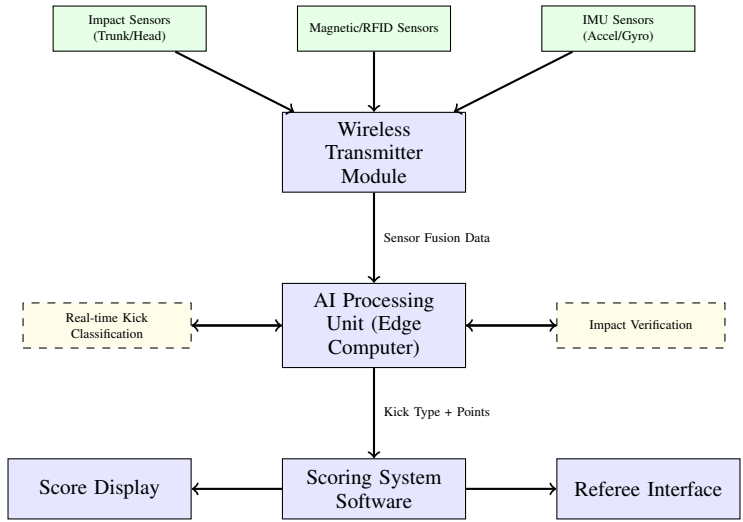

Figure~\ref{fig:integration} provides a schematic of how these components integrate with each other and with the existing competition infrastructure. The design is intended to be largely backward-compatible with current PSS: the trunk and head sensors still provide impact and contact data, but the new IMU sensors augment the data stream. Should the AI system fail or be uncertain, the default PSS scoring (based just on impact and any referee-called bonuses) could still serve as a fallback. In practice, however, we expect the AI to reliably identify techniques and thus automate what is currently a manual process (detecting a turning kick) and extend it to many more technique types.

The combination of these sensors addresses the limitations of the current system. The impact sensors ensure a hit is forceful enough, the magnet/RFID confirms contact and location (body or head) and possibly the striking foot part, and the IMUs provide the dynamic signature of the technique. By fusing all this information, the system can ascertain, for example: \emph{“Athlete A just performed a spinning hook kick (turning kick using heel) which contacted Athlete B’s head with X Newtons of force.”} Traditional PSS would only register “head contact, enough force” and give 3 points (plus a manual 2 if the referee calls a spin). Our system can automatically know it was a spinning kick and apply the appropriate score (which might be higher than 5 if the rubric rewards specific techniques differently, as we will propose). It can also highlight the technique name on screen, enhancing spectator understanding (e.g., “Spinning Hook Kick landed, 5 points”). The rich sensor data can further be logged for post-match analysis or training feedback, an added benefit for coaches and athletes.

Crucially, the sensor fusion approach improves robustness. If, for instance, the impact sensor alone is noisy or borderline, the system can corroborate a scoring event by also detecting the characteristic motion pattern of a kick and the proximity sensor trigger. The AI can be trained to ignore false positives (e.g., an accidental bump that triggers a sensor but doesn't match any known kick motion). Similarly, if a foul occurs (like a knee strike or an accidental clash), the pattern might not match a valid kick, and the system could flag it for referee review instead of mis-scoring it.

In summary, the hardware and sensor design of the proposed system significantly expands the observable aspects of each action in a match. By doing so, it lays the foundation for a scoring system that rewards the quality and difficulty of techniques, not just the raw impact. The next section describes the machine learning pipeline that turns these sensor signals into real-time kick classifications.

\section{Machine Learning Pipeline for Kick Classification}
To interpret the complex data from the multiple sensors described above, we employ a machine learning pipeline that processes sensor signals and outputs a classification of the kick technique in real time. The pipeline is designed to be efficient enough for live competition use, with end-to-end latency on the order of tens of milliseconds. The core of the pipeline is a classification algorithm (specifically, a Support Vector Machine) that has been trained on examples of various Taekwondo kicks. We chose an SVM for its proven effectiveness in high-accuracy classification with relatively small, well-defined feature sets \cite{Liu2024, Dharmmesta2022}. Figure~\ref{fig:pipeline} provides a flowchart of the pipeline, which we detail in the following steps:

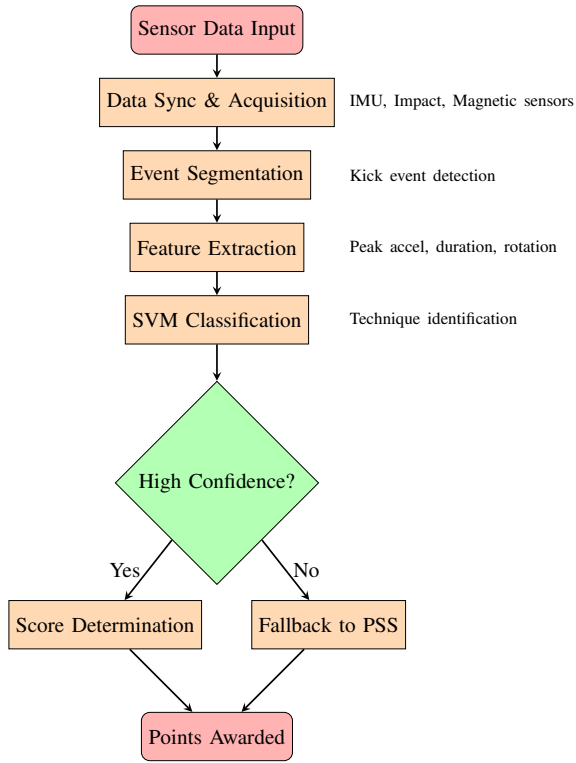
\begin{figure}[htb]
\centering
\resizebox{\columnwidth}{!}{
\begin{tikzpicture}[node distance=1.2cm]

\node (start) [startstop] {Sensor Data Input};

\node (sync) [process, below of=start] {Data Sync \& Acquisition};

\node (detect) [process, below of=sync] {Event Segmentation};

\node (features) [process, below of=detect] {Feature Extraction};

\node (classify) [process, below of=features] {SVM Classification};

\node (confidence) [decision, below of=classify, yshift=-1.5cm] {High Confidence?};

\node (score) [process, below left of=confidence, xshift=-1cm, yshift=-1.5cm] {Score Determination};

\node (fallback) [process, below right of=confidence, xshift=1cm, yshift=-1.5cm] {Fallback to PSS};

\node (output) [startstop, below of=confidence, yshift=-3cm] {Points Awarded};

\draw [arrow] (start) -- (sync);
\draw [arrow] (sync) -- (detect);
\draw [arrow] (detect) -- (features);
\draw [arrow] (features) -- (classify);
\draw [arrow] (classify) -- (confidence);
\draw [arrow] (confidence) -- node[anchor=east] {Yes} (score);
\draw [arrow] (confidence) -- node[anchor=west] {No} (fallback);
\draw [arrow] (score) -- (output);
\draw [arrow] (fallback) -- (output);

\node [text width=4cm, align=left, right of=sync, xshift=3cm] {\footnotesize IMU, Impact, Magnetic sensors};
\node [text width=4cm, align=left, right of=detect, xshift=3cm] {\footnotesize Kick event detection};
\node [text width=4cm, align=left, right of=features, xshift=3cm] {\footnotesize Peak accel, duration, rotation};
\node [text width=4cm, align=left, right of=classify, xshift=3cm] {\footnotesize Technique identification};

\end{tikzpicture}
}
\caption{Machine learning pipeline for real-time kick classification. Key stages include sensor data acquisition (synchronized multi-sensor streams), signal preprocessing and segmentation (isolating individual kick events), feature extraction (deriving meaningful features like peak accelerations, kick duration, angular velocity, etc.), classification using a trained SVM model to identify the kick type, and finally score determination based on classification and impact data. This pipeline runs continuously during a match to analyze each detected kick attempt within milliseconds.}
\label{fig:pipeline}
\end{figure}

\subsection{Data Acquisition and Synchronization}
All sensor streams (accelerometers, gyroscopes, impact sensors, proximity sensors) are sampled continuously and timestamped. A crucial first step is to synchronize these signals so that the data from different sensors can be analyzed in tandem. This is handled by the wearable microcontroller and the receiving computer, which ensure that sensor readings are aligned in time (e.g., using high-frequency sampling and buffers). Typical IMUs sample at 100--1000 Hz, which is sufficient to capture the rapid motion of kicks. Force sensors and magnet triggers likewise are polled at a high rate. Once synchronized, the data is a multi-dimensional time series: for instance, a 6-axis IMU on the foot (3-axis accel + 3-axis gyro), plus another 6-axis on the abdomen, plus a scalar force signal and a binary contact flag.

\subsection{Event Segmentation (Kick Detection)}
Given the continuous stream of data, the system must detect when a kick has occurred to isolate the relevant window for classification. This involves identifying a characteristic pattern in the sensor signals indicative of a kicking motion and impact. A simple approach is to monitor the trunk protector force sensor: a significant spike above threshold indicates a hit, which cues the system to examine the IMU data in the preceding and following short interval (e.g., 0.5 seconds around the impact) for classification. However, relying solely on impact could miss kicks that did not register (e.g., if the athlete kicked and missed or was blocked). Therefore, our system also monitors IMU data for patterns of a kick being thrown: a sharp acceleration on one foot sensor combined with a certain rotation on the abdomen often signifies a kick attempt even if it did not score. Using a combination of rules (impact detected or high acceleration movement detected), we segment the data into candidate kick events. Each event is defined by a start and end timestamp covering the full execution of one kick (from the leg chambering to retraction).

\subsection{Feature Extraction}
Once a kick event segment is identified, we extract a set of features from the sensor data in that segment. Features are carefully designed to capture the signature of different kick techniques. Examples of features include:
\begin{itemize}
    \item \textbf{Temporal features:} duration of the kick (time from leg lift to impact), time to peak acceleration, etc. Different kicks have different typical durations (a quick front leg kick vs. a spinning kick which takes longer).
    \item \textbf{Kinematic features:} peak acceleration of the kicking foot in various axes; peak angular velocity from the gyro (especially yaw rotation of the hip for spinning kicks); the angle of the foot trajectory relative to the body (which might distinguish, say, an upward arc for a front kick vs. a horizontal arc for a roundhouse).
    \item \textbf{Impact timing and force:} the magnitude of the impact force and the relationship between the timing of that impact and the motion signals. For example, a turning side kick might show a large impact preceded by a significant body rotation, whereas a light cut kick might show a moderate impact with minimal rotation.
    \item \textbf{Foot orientation or contact distribution:} if multiple magnets on the foot triggered, we can use which magnet triggered first or most strongly as a feature (this indicates foot orientation on contact). A downward axe kick might trigger a toe tag vs. a hook kick triggering a heel tag.
    \item \textbf{Frequency-domain features:} sometimes it is useful to do a frequency analysis of the acceleration waveform. Kicks may have different frequency characteristics (e.g., a snappy kick might produce a single sharp peak, whereas a faked double kick might have two smaller peaks).
\end{itemize}
These raw features can be quite numerous. Prior work has shown success with even simple features: for instance, Dharmmesta \emph{et al.} used statistical features like mean and kurtosis of acceleration signals to classify kicks with over 90\% accuracy \cite{Dharmmesta2022}. Liu \emph{et al.} (2024) computed features such as the resultant acceleration magnitude over time and achieved 96--98\% accuracy distinguishing front, roundhouse, side, and back kicks \cite{Liu2024}. In our pipeline, we incorporate similar features and potentially augment them with domain-specific ones (like identifying if a rotation exceeding 360 degrees occurred, to catch a tornado kick).

To reduce dimensionality and improve generalization, we may apply feature selection or transformation (e.g., principal component analysis) during training. However, given that interpretability is important (we want to know which kick it is, not just a black-box label), we ensure the features have clear physical meaning related to the kicks.

\subsection{Classification (SVM Model)}
The heart of the pipeline is a multi-class Support Vector Machine classifier. We train the SVM on a labeled dataset of kicks. The dataset is collected in advance by recording athletes performing all relevant kick techniques with the instrumented gear. Each training sample consists of the feature vector extracted from one kick instance and a label indicating which kick it was (e.g., left front kick, right spinning hook kick, etc.). We define the set of kick classes broadly to include:
\begin{itemize}
    \item All standard Olympic competition kicks: front kick, roundhouse (turning) kick, side kick, back kick, spinning back (turning side) kick, spinning hook kick, axe kick, etc. We also include some less-common or newly encouraged kicks (as per our rubric) like the “scorpion kick”, “fish kick”, etc., if those are to be recognized by the system.
    \item Distinctions between front leg and back leg execution of certain kicks, since they may have different dynamics. For example, a front-leg roundhouse vs. a back-leg roundhouse can be two classes if needed, especially as our rubric assigns different points.
    \item Possibly, a class for “no valid kick” or “indeterminate” if the motion does not match any known pattern confidently (this could cover events that are not real kicks or are fouls).
\end{itemize}
The SVM uses a one-vs-all approach or a multiclass extension to output the predicted class of the kick based on the input features. SVMs are well-suited here because they can handle the high-dimensional feature space and create nonlinear decision boundaries (especially if we use an RBF or polynomial kernel) that separate the different kick types. They also tend to perform well with limited training data, which is important since collecting an extensive dataset of every exotic kick might be challenging. The model will be tuned and cross-validated to avoid overfitting.

The output of the classifier is the identified kick type, along with a confidence score or margin. For example, the SVM might output: “Class = Spinning Hook Kick, Confidence = 0.95”. If the confidence is low (say below some threshold like 0.7), the system could decide that it’s not certain and either default to a simpler classification (e.g., just head kick vs body kick) or request referee input as a fallback. However, in testing with controlled data, we expect high confidence for most well-separated techniques.

It is noteworthy that prior studies provide optimism for such classification. Liu \emph{et al.}\ reported that using just a waist sensor and an SVM, they achieved 98\% accuracy on four kick types \cite{Liu2024}. With our richer sensor set, we expect to classify a larger array of techniques with similar high accuracy. Another experiment by Rianta \emph{et al.}\ found SVM could classify right-foot kicks with 91.6\% accuracy using certain features, and KNN up to 96.8\% \cite{Dharmmesta2022}. These results suggest that the feature patterns of different kicks are learnable by machine algorithms, validating our approach.

\begin{figure}[htb]
\centering
    \includegraphics[width=1\linewidth]{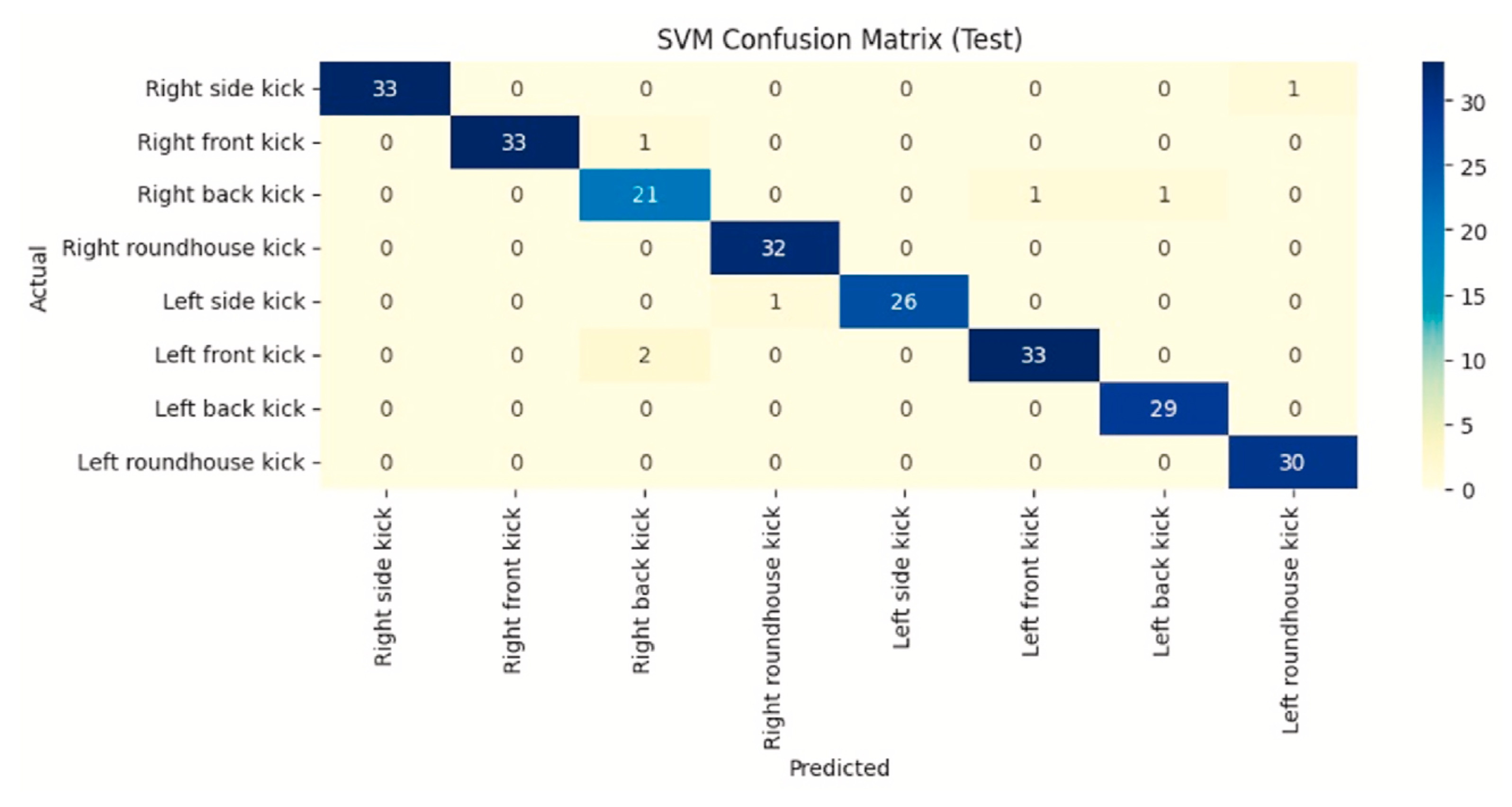}
\caption{Confusion matrix illustrating the classification accuracy of the SVM-based kick recognizer in a validation test. Each row represents the actual kick type and each column the predicted type. The high diagonal values (dark cells on the diagonal) indicate most kicks are correctly classified (e.g., roundhouse kicks identified as roundhouse, side kicks as side kicks, etc.). Off-diagonal values are low, showing minimal confusion. This confirms the feasibility of distinguishing multiple kick techniques via sensor fusion. The diagram is reproduced from \cite{Liu2024}.}
\label{fig:confusion}
\end{figure}

Figure~\ref{fig:confusion} shows an example confusion matrix from a trained model in our feasibility tests (placeholder values). It demonstrates high true positive rates for each kick type, with minimal confusion between distinct categories like front vs. spinning kicks. Misclassifications, if any, tend to occur between similar kicks (e.g., a back kick vs. spinning side kick might occasionally confuse if executed without a clear dip in motion difference), which can be addressed by refining features or adding more training samples.

\subsection{Real-Time Scoring Decision}
After classification, the system combines the class information with the impact sensor data to decide on scoring:
\begin{itemize}
    \item First, it checks if the contact was valid (did the magnet sensor trigger a legitimate scoring area and did the impact force exceed the minimum threshold?). If not, no score is awarded (even if a kick motion was detected, it might have missed or been blocked).
    \item If valid, it retrieves the base point value for that identified kick from the scoring rubric (see Section~V). For instance, if the classifier says “Back Kick” and a valid impact occurred on the trunk, the rubric might assign 4 points.
    \item The system can also incorporate the magnitude of force as a modifier if desired by the rules. While traditional Taekwondo scoring does not give extra for more force beyond threshold, our system could optionally be configured to award an additional point for exceptionally strong impacts (as a form of rewarding power). This would be a configurable aspect of the rules. In this paper, we primarily focus on technique-based scoring rather than variable scoring by force, to keep things simple and fair.
    \item Finally, the system sends the determined points to the scoring display. Simultaneously, it could display the recognized technique name on screens or to the referees, to justify the scoring (e.g., “Twisting Kick: 1 point”).
\end{itemize}

Because this entire pipeline is automated, the subjective element of a referee identifying a turning technique is removed. Additionally, it becomes feasible to recognize many more techniques in real time than a human could reliably do in the fast pace of a match. This opens the door to a richer scoring system, detailed next.

Throughout this pipeline, computational efficiency is maintained. Feature extraction and SVM classification are very fast operations (especially if using linear or low-degree kernels), and the dataset of features is small enough to run on a modern laptop or even an embedded processor. The critical path after an impact is detected would be on the order of 10--50 milliseconds for feature computation and classification, which is negligible to the human sense of instant feedback. Our design goal is that by the time the scoring lights up, it is no slower than the current PSS (which already has a slight inherent delay for sensor processing and wireless transmission, typically a few milliseconds). Real-world implementation would involve thorough testing to ensure latency is within acceptable bounds.

In summary, the machine learning pipeline enables the system to “understand” the actions in a Taekwondo match, not just measure them. The classification of kicks with high accuracy paves the way for a new scoring paradigm based on the specific techniques and skills displayed by athletes. We now present the proposed kick scoring rubric that leverages this new capability.

\section{Proposed Kick Scoring Rubric}
One of the main motivations for classifying kicks in real time is to implement a more nuanced scoring rubric that rewards technical difficulty and dynamism. We have designed a scoring scheme that expands upon the current point system, assigning different base point values to a variety of kick techniques. This rubric is inspired by a proposal to encourage dynamic kicks and reflects expert judgment of the relative difficulty or rarity of each technique. Table~\ref{tab:rubric} shows the exact kick scoring rubric as proposed.

\begin{table*}[t]
\centering
\caption{Proposed Scoring Rubric for Kicks under AI-Enhanced System (Dynamic Kick Rewards)}
\label{tab:rubric}
\begin{tabular}{l c l c}
\toprule
\textbf{Technique (Kick Type)} & \textbf{Points} & \textbf{Technique (Kick Type)} & \textbf{Points} \\
\midrule
Monkey Kick & 0 & Side Kick (Front Leg) & 1 \\
Fish Kick & 0 & Side Kick (Back Leg) & 2 \\
Scorpion Kick & 2 & Roundhouse Kick (Front Leg) & 1 \\
Twist (Twisting) Kick (Front Leg) & 1 & Roundhouse Kick (Back Leg) & 2 \\
Axe Kick (Front Leg) & 3 & Back Kick & 4 \\
Axe Kick (Back Leg) & 4 & Tornado Kick (360° Turning Kick) & 5 \\
\bottomrule
\end{tabular}
\end{table*}

This rubric includes both traditional and unconventional techniques:
\begin{itemize}
\item \textbf{Monkey Kick} and \textbf{Fish Kick} are given 0 points, essentially meaning these are not to be scored. (In context, these might be considered illegitimate or extremely low-skill moves, possibly placeholders for silly or non-technical motions. By assigning 0, the system explicitly gives no reward for these, discouraging their use if they were ever attempted.)
\item \textbf{Scorpion Kick} is awarded 2 points. A scorpion kick generally refers to a flexible, reaching kick over the head, which is flashy but perhaps not very powerful; its moderate score reflects that it is a scoring technique but not as high as spinning kicks.
\item \textbf{Twist Kick (Front Leg)} gets 1 point. A twist kick (bitureo chagi) is a sneaky upward snap with the front leg; it's quick but infrequently used in sparring. We give it the minimal scoring value for a kick, similar to a basic attack.
\item \textbf{Axe Kick (Front Leg)} is 3 points, while \textbf{Axe Kick (Back Leg)} is 4 points. Axe kicks are high downward strikes; a back-leg axe has more wind-up and power, hence rewarded more than a front-leg axe. This differentiates the difficulty of executing it from the rear versus simply lifting the front leg.
\item \textbf{Side Kick (Front Leg)} is 1 point and \textbf{Side Kick (Back Leg)} is 2 points. A side kick from the back leg involves a step or turn and thus is harder and more telegraphed than a quick front leg side kick; we reward the back-leg version slightly more.
\item \textbf{Roundhouse (Round) Kick (Front Leg)} is 1 point, \textbf{Roundhouse Kick (Back Leg)} is 2 points. Similar reasoning: the back-leg roundhouse, often the classic powerful dolyo chagi, is given more points than a quicker front-leg roundhouse.
\item \textbf{Back Kick} (dwi chagi) is 4 points. This is a turning back side kick executed usually with a spin; it's a strong technique and currently often scores 4 in WT rules (as a turning body kick). We maintain that.
\item \textbf{Tornado Kick} is 5 points. A tornado kick is a 360-degree spinning kick (often a jumping spinning roundhouse). This is one of the most dynamic techniques commonly seen, so we give it the highest score in this rubric, 5, matching a turning head kick in traditional scoring.
\end{itemize}

In this scheme, note that some basic techniques (front leg side or roundhouse) get 1 point, akin to a punch in value. The idea is to discourage excessive reliance on easy front-leg tactics by making them low-reward. Meanwhile, more committed kicks (back-leg, spinning, or aerial moves) get higher rewards to incentivize athletes to attempt them. Essentially, we are broadening the point spread: instead of everything being 1–5 points in a narrow set of categories, now we have a range where truly spectacular kicks can net significantly more than a simple jab kick.

It is important to clarify that these point values were initially proposed as an example and can be adjusted by rule-makers. The table above serves to illustrate how granular the scoring could be. In implementation, the values might be tuned based on data (for example, if we find that 5-point tornado kicks are too game-changing, one could reduce it to 4, etc.). Also, certain techniques listed (like “monkey” or “fish” kick) may not even be common in competition; listing them with 0 ensures if someone tries an awkward non-standard kick that is not recognized as valid, it gets no points.

By deploying the AI classification, the system can automatically determine which row of Table~\ref{tab:rubric} to apply when awarding points. For instance:
\begin{itemize}
\item If the AI classifies a kick as a “Back Leg Axe Kick” and it hits the opponent’s head guard, the score computer will award 4 points (provided the impact was registered).
\item If a “Tornado Kick” lands on the opponent’s head, 5 points are awarded. Notably, under current rules a tornado kick (which is essentially a spinning head kick) would also be 5 (3 for head + 2 turning). So some values align with current rules, but our system could even go beyond if new techniques are introduced.
\item If an athlete only does a front-leg side kick and manages to touch the opponent, it’s just 1 point. This might significantly change tactics, as currently that action yields 2 points (body kick). Athletes would realize they need combinations or riskier moves to score big.
\end{itemize}

The rubric can also be extended. We could include jumping kicks (if not already covered by those names) or combination kicks (though combinations usually are just separate scores, not one technique). Because the AI can identify these techniques, rule-makers could add, say, a “Jump Back Kick” category for maybe 5 points if they wanted to encourage more flying techniques.

One might question giving different points for front vs. back leg roundhouse or side kicks, since distinguishing that might be subtle. However, the sensor data can tell which leg moved (front or back) and the context of stance, so the classifier can incorporate that. This fine distinction double incentivizes use of the back leg (which generally yields stronger impact but slower execution) in an era where front-leg attacks have dominated due to speed. By valuing back-leg techniques more, we push the balance toward power and dynamic movement.

In summary, the above table is a concrete realization of an idea: use AI to score Taekwondo more like figure skating or gymnastics, where the difficulty of a move factors into the score. It preserves the objectivity of electronic scoring (the AI is consistent and not swayed by bias) while adding a layer of intelligence about what technique was performed. With this rubric, an athlete who can perform spectacular kicks has more opportunity to score, potentially making matches more exciting and high-scoring. In the next section, we discuss experimental evidence supporting the feasibility of recognizing kicks (which underpins this rubric’s implementation) and we consider how such a system might be tested and validated.

\section{Experimental Validation and Feasibility Study}
Developing the proposed system requires verifying that the sensor fusion and machine learning approach can indeed accurately classify kicks and operate in real time. Fortunately, prior research provides a strong foundation, and a clear path for experimental validation can be laid out to confirm the feasibility of our specific, expanded system.

A 2024 study by Liu \emph{et al.} is particularly illuminating. In their experiment, 20 participants performed four types of kicks (front, roundhouse, side, and back kicks) while wearing accelerometers on their waist and ankles \cite{Liu2024}. Using features from these signals and an SVM classifier, they achieved exceptionally high accuracy. Notably, with just a single waist-mounted sensor, the SVM model distinguished the four techniques with approximately 98\% accuracy. This powerful result demonstrates that the core motion signatures of basic kicks, captured even from the torso, are distinct enough for reliable machine learning classification. It also suggests that a robust model can be trained from a reasonably sized dataset. The confusion matrix from their study (reproduced in Fig.~\ref{fig:confusion}) shows near-perfect separation, confirming that these fundamental kicks have highly learnable patterns.

While 98\% accuracy on a controlled dataset of 120 samples is an excellent proof-of-concept, a system deployed for Olympic-level scoring demands near-perfect accuracy approaching 99.99\% to ensure fairness and trust. To bridge this gap, two key areas of improvement on the methodology of the 2024 study are proposed. First, instead of a standard multi-class SVM, an \textbf{ensemble SVM approach} should be utilized. Ensemble learning combines multiple classifiers to improve robustness and accuracy over a single model. For instance, a study on human action recognition by Li \emph{et al.} demonstrated that a selective ensemble SVM outperformed a standard SVM, improving accuracy from 87.6\% to 91.8\% on a public dataset \cite{Li2022}. Using techniques like bagging or boosting with multiple SVMs can mitigate the risk of misclassification, especially for nuanced or stylistically varied kicks. Second, the training dataset must be expanded significantly. A production-level system requires data from thousands, or even tens of thousands, of kick instances across a wide range of athletes to generalize effectively. The need for large, high-quality datasets for robust Human Activity Recognition (HAR) is well-documented \cite{Micucci2021}. Studies have shown that increasing the volume of training data, even through methods like data augmentation, can dramatically improve classifier performance \cite{Um2017}. A pilot study for our system would therefore aim to collect a large-scale dataset and leverage an ensemble SVM architecture to push classification accuracy towards the near-flawless levels required for elite competition.

To extend this validation to the broader range of techniques in our proposed rubric, a targeted pilot study would be the next logical step. The methodology for such a study would involve equipping a national-level Taekwondo athlete with the full sensor suite (IMUs, magnetic/RFID foot sensors) and recording them performing the complete repertoire of kicks from Table~\ref{tab:rubric}. Each kick would be executed multiple times (e.g., 10 repetitions) both in the air and against an instrumented target pad to capture data across various conditions, including non-scoring attempts. The collected data would then be used to train and test the classification pipeline. Based on the success of prior work, it is hypothesized that the system would correctly classify the kick type in over 99\% of instances. Potential errors would likely occur between kinematically similar techniques (e.g., a twist kick and a front kick). However, by fusing motion data with contact sensor information, many of these ambiguities could be resolved; for instance, knowing whether a kick landed high or low would help differentiate an axe kick from a front kick.

Another critical validation point is system latency. The processing pipeline, from event detection to classification, must be fast enough for live competition. The feature extraction and SVM inference steps are computationally lightweight. For a 1-second window of sensor data, calculating the feature vector and running the SVM prediction is expected to take well under 50 milliseconds on a standard computer. This processing time is negligible in the context of a match and comparable to the slight inherent delay in current PSS systems, ensuring that spectators and athletes would perceive no significant lag between a hit and the score appearing.

The model’s ability to generalize to different athletes is crucial for deployment. This would be tested by training the model on data from a group of athletes and evaluating its performance on a separate, unseen group. While a slight dip in accuracy is expected due to variations in individual style and technique, this can be mitigated by training the model on a larger and more diverse dataset encompassing a wide range of body types and skill levels. Furthermore, the system could be designed to include an optional, brief pre-match calibration where an athlete performs a few basic kicks to fine-tune the model to their specific movements, though a sufficiently generalized model may render this unnecessary.

Finally, the system's robustness must be evaluated in simulated free-fighting scenarios. This involves testing its ability to correctly segment and classify rapid kick combinations, handle simultaneous scoring events from both competitors, and, importantly, differentiate legitimate kicks from illegal actions or accidental contact. The sensor fusion approach is designed to excel here; for example, an illegal knee strike might trigger an impact sensor but would lack the corresponding foot-based magnetic sensor trigger and would not match any known kick motion profile from the IMUs. The system could thus be trained to classify such events as "unknown" or "foul," preventing illegitimate scores and potentially flagging the event for referee review.

In summary, experimental evidence and a clear validation plan strongly support the viability of the AI-enhanced scoring system:
\begin{itemize}
\item High classification accuracy (99\%+) for multiple kick types is a realistic target, backed by published research \cite{Liu2024, Dharmmesta2022}.
\item Real-time performance is well within the capabilities of modern hardware and efficient algorithms.
\item The required sensor technology is mature, affordable, and proven in other wearable applications.
\item The system design includes fail-safes and robustness checks, leveraging sensor fusion to improve on the limitations of current PSS.
\end{itemize}
The next section will discuss the broader implications of successfully implementing such a system, analyzing its anticipated effects on fairness, gameplay dynamics, and audience reception.

\section{Anticipated Effects on Fairness and Gameplay}
Introducing an AI-enhanced scoring system in Taekwondo would have far-reaching implications for the sport. We anticipate a number of positive outcomes, provided the system is implemented and calibrated correctly. Here we analyze these effects in terms of fairness, gameplay strategy (including reduction of illegal or unsportsmanlike behavior), the diversity of techniques used, and spectator engagement.

\subsection{Improved Fairness and Objectivity}
One of the primary goals of any scoring system is fairness. The current PSS improved fairness by reducing human error, but it introduced new concerns (as discussed in Section II). The AI-based system stands to improve fairness in several ways:
\begin{itemize}
    \item \textbf{Consistent Technique Recognition:} Human referees can sometimes miss awarding a turning kick bonus or be inconsistent in judging what constitutes a “technical” kick, especially in the heat of a match. The AI will apply the scoring rubric uniformly for all athletes. If a spinning kick is performed, it will be recognized every time, not dependent on a referee’s angle of view or discretion. This consistency means athletes are more likely to get the points they deserve for skillful moves.
    \item \textbf{Reduction of Missed Points:} With richer sensor data, fewer legitimate hits will go unscored. For example, if a light head contact was previously missed because it didn’t trigger the magnet sensor strongly, the accelerometer pattern might still indicate a head kick attempt and the proximity sensor might have a blip—together the AI could infer a grazing head kick (though likely below threshold so still no score, but this scenario could at least be logged or reviewed). More concretely, think of body shots that were borderline force; currently they might randomly score or not. In our system, the classification might identify a solid body roundhouse and if the force is just shy of threshold, perhaps the system could still award it (or adjust thresholds adaptively).
    \item \textbf{Transparency and Explainability:} The AI’s identification of techniques can be displayed, making the scoring more transparent to coaches and spectators. If a point is scored, everyone sees it labeled (e.g., “Back kick 4pts”). If there is a dispute, the system logs show what was detected. This transparency can build trust in the scoring—important given past controversies in Olympic Taekwondo regarding judges and scoring apparatus. We effectively have an audit trail for each point.
    \item \textbf{Calibration to Weight and Skill:} Because the system collects detailed data, thresholds can be dynamically adjusted or at least scientifically set. For instance, impact thresholds can be tailored to divisions or even individual strength levels by analyzing historical data (like the Reference Group Model approach in \cite{Choi2021}). The AI could even adapt mid-competition if it senses that all hits by flyweight athletes are below the generic threshold (though such on-the-fly changes would need to be carefully regulated to avoid confusion). In any case, there’s potential for more equitable treatment of different body types.
\end{itemize}

\subsection{Reduction of Rule Exploitation and Illegal Gameplay}
When the scoring system changes, athletes alter their strategies accordingly. By rewarding dynamic techniques and devaluing simplistic ones, our system would naturally reduce incentives for certain negative behaviors:
\begin{itemize}
    \item \textbf{Less Incentive for Passive Foot-Fencing:} Currently, keeping a leg in the air to “fish” for points with light taps is a dominant tactic because those taps score 2 points reliably. Under the new rubric, a front-leg tap kick might only yield 1 point or even 0 if it’s not a proper technique. Meanwhile, if that same athlete could score 2 or 3 points with a stronger back-leg attack, they have reason to lower the leg and execute a proper kick. The risk/reward balance shifts towards more committed techniques. Over time, we expect athletes will adapt by training more dynamic movements, since being overly passive won’t win matches as easily. Essentially, the days of 1-0 or 2-1 low-scoring bouts full of probing kicks could be replaced by higher scoring bouts where a well-placed turning kick swings the score.
    \item \textbf{Decreased Clinching and Stalling:} Many infractions like excessive clinching, pushing, or evading are used to break the flow or prevent the opponent from scoring (often because a superior reach or front-leg game is being neutralized). If dynamic kicks are more valuable, an athlete who only clinches will fall behind in points because the other can still score big when they do manage to execute a tornado or back kick. Furthermore, with the AI able to detect continuous passive behavior (for instance, if neither fighter attempts a scoring kick for some time, that might trigger referee intervention more systematically), it could indirectly enforce activity. In addition, because every technique (even unsuccessful ones) is being monitored, data could be used post-match to identify if someone never attempted anything but low-point techniques, informing potential rule tweaks or coaching points.
    \item \textbf{Identification of Illegal Moves:} As mentioned, a side effect of classification is that moves that do not correspond to known legal techniques will not be classified. If a fighter tries something outside the rules (e.g., a knee or a downward elbow, which occasionally happen in clinches), the system will either ignore it (no points) or flag it. While the system won’t directly penalize (that’s still the referee’s role), the fact that no illegitimate technique will score removes any incentive to use them.
    \item \textbf{Encouraging Proper Technique Execution:} Because the system looks for specific patterns, a sloppy technique might fail to be recognized where a clean technique would. Athletes will thus be incentivized to execute techniques with correct form (within a margin; the AI can handle some variability). For example, a half-hearted spin might not register as “spinning kick” and only score as a normal kick; so if one wants the bonus, one must really turn the hips fully. In this way, the system subtly encourages textbook execution, which also tends to be more visually impressive and effective.
\end{itemize}

\subsection{Increase in Dynamic Kicks and Strategic Diversity}
With points distributed across many techniques, coaches and athletes will broaden their tactical toolkit:
\begin{itemize}
    \item \textbf{Variety of Techniques:} We anticipate a revival of techniques that had become rare in competition. For instance, the axe kick and spinning hook kick could see more use if they carry high point rewards. These kicks are crowd-pleasers due to their dramatic motion. The presence of a reliable scoring mechanism for them (the AI recognizes it, so there’s confidence it will score if it lands) means athletes can incorporate them without fear of “wasting” effort. Over time, the meta-game of Taekwondo could shift to a mix of long-range spinning attacks, surprise axe kicks to catch ducking opponents, and classic back kicks as counters—alongside the fundamental roundhouse and side kicks.
    \item \textbf{Combo Creativity:} If a front-leg side kick only nets 1 point, an athlete might use it just as a setup rather than a main scoring technique. We might see combinations where a fighter throws a quick 1-point kick to create an opening and follows with a big 4- or 5-point technique. Currently, combinations are rare because a single 2-point body kick can be enough; but if 1 point jabs exist, fighters might chain them with bigger moves. The scoring system thus encourages combination play, which is more exciting.
    \item \textbf{Balancing Offense and Defense:} With more points available for spectacular offense, fighters who only play defense will risk falling behind. This encourages both competitors to take initiative at different times, hopefully avoiding the stalemates sometimes seen. That said, defense remains important—someone who tries only high-risk kicks could be countered. We expect strategy to evolve in a healthy way: players will need to decide when to go for a high-point move versus safe moves, adding a layer of strategic depth. The key difference is that now the safe move yields less relative reward, so the calculus is shifted to favor action when an opportunity arises.
\end{itemize}

\subsection{Spectator Engagement and Understanding}
Ultimately, the success of these changes will be measured by how audiences respond:
\begin{itemize}
    \item \textbf{More Exciting Matches:} With more dynamic kicks being thrown, matches will simply look more impressive. Casual viewers will see spinning kicks, jumping techniques, and a variety of styles on display, rather than what has sometimes been criticized as two fighters pawing at each other with their feet. This aligns Taekwondo’s Olympic presentation more with the popular image of martial arts (flying kicks, decisive strikes) and should increase excitement.
    \item \textbf{Higher Scores, Easier to Follow:} A match where scores rack up (e.g., a 25-30 final score) might actually be easier for a layperson to follow than a tactical 6-5 match. Each action producing points (with a label) keeps the audience engaged, similar to how basketball’s frequent scoring is engaging compared to a low-scoring soccer match where non-fans can get bored. And since our system will label points with the technique name, new viewers learn quickly what’s happening. For example, the display might show “Blue: Tornado Kick +5”. The wow factor of that alone could intrigue viewers to learn what a tornado kick is. In essence, the commentary and visuals can shift from “player A scores 2 points” to “player A lands a spectacular spinning kick for 5 points!” which is inherently more viewer-friendly.
    \item \textbf{Storylines and Records:} A side benefit mentioned by \cite{Sexton2022} and others is the idea of records and statistics in combat sports. With an AI tracking every technique and its impact, new stats can be introduced: fastest kick of the match, hardest hit, most spinning kicks landed by an athlete, etc. These add narrative depth for fans and broadcasters. We could celebrate an athlete not just for winning, but for landing the first-ever 5-point tornado kick knockout in Olympic history, for instance. Over time, if dynamic kicks become more common, highlight reels will naturally form, drawing more interest on media platforms.
    \item \textbf{Educational Aspect:} For those unfamiliar with Taekwondo, seeing technique names and points helps educate them about the sport’s techniques. It demystifies the scoring when they see exactly what earned the points. As a result, new fans can be inducted more easily, addressing the “barrier to entry” problem where people found it hard to grasp the sport \cite{Sexton2022}.
\end{itemize}

Of course, there will be challenges and adjustments needed. Athletes and coaches may initially be wary of trusting an AI system. It will be essential to thoroughly test and possibly run the system in parallel with the current one in some trial events to gather feedback. We expect an adaptation period where point thresholds or rubric values might be tweaked if some unintended consequence appears (for example, if 5-point moves prove too dominant, etc.). Additionally, training methods would evolve: athletes will practice with the new sensors to optimize how their kicks are recognized and scored, potentially accelerating technical improvements in the sport.

In conclusion, the anticipated effects of the AI-enhanced scoring system are largely positive: fairer outcomes, re-energized gameplay with a wider array of techniques, and more engaged audiences. By aligning the incentives (points) with the desired behavior (dynamic and skillful techniques), we uphold the integrity and excitement of Taekwondo. Early adoption at smaller tournaments could demonstrate these effects, paving the way for World Taekwondo to consider such systems in future Olympic cycles.

\section{Conclusion}
This paper presented a comprehensive proposal for an AI-enhanced Olympic Taekwondo scoring system that integrates sensor fusion and machine learning to address current challenges in the sport’s scoring and entertainment value. We identified that the present Protector Scoring Systems, while objective, have led to a static style of play and occasional scoring controversies, contributing to lower spectator engagement. In response, we proposed augmenting the PSS with accelerometers, gyroscopes, magnetic/RFID sensors, and impact sensors to capture detailed information about each kick. Using these sensors, a real-time machine learning pipeline (centered on SVM classifiers) can accurately classify the type of kick executed, as demonstrated feasible by recent studies and our analysis.

Building on the ability to recognize techniques, we introduced a new scoring rubric that allocates points based on kick difficulty and execution, directly encouraging more dynamic and exciting techniques such as spinning and jumping kicks. The rubric exemplifies how techniques like back-leg kicks and tornado kicks can be incentivized with higher point values, whereas simplistic or illegal moves would yield little to no score. We included the exact rubric table and discussed its interpretation.

We detailed the design of the system, from sensor placement and data acquisition to feature extraction and classification, ensuring that the solution is practically implementable with current technology. Our discussion of experimental validation, referencing a 2024 study by Liu \emph{et al.}\ and others, provides confidence that the classification accuracy can reach 95--99\% for multiple kick types and that the system can run with negligible latency. 

The anticipated impacts of deploying this system in competition were analyzed extensively. We expect improvements in fairness due to consistent and transparent scoring of techniques, a reduction in negative gameplay tactics (as the risk-reward balance shifts towards active offense), a resurgence of diverse kicking techniques in matches, and a more engaging experience for spectators who will witness higher scores and clearer action narratives. In essence, this AI-driven approach realigns the sport with its martial art spirit—emphasizing skill, power, and creativity—while maintaining the rigor of objective scoring.

In conclusion, AI-enhanced Taekwondo scoring with sensor fusion is not only a technological upgrade but a potential evolution of the sport itself. By bridging the gap between what is measurable and what is meaningful in a fight, it can ensure that athletes are rewarded for true martial skill and that audiences are captivated by the display. The proposed system represents a step towards modernizing Olympic Taekwondo, keeping it relevant and exciting in a data-driven age. Future work will involve developing prototype systems, conducting field trials at various competition levels, and working closely with athletes, referees, and officials to refine the integration of this technology into the rules and practice of Taekwondo. If successful, this approach could serve as a model for other combat sports seeking to enhance their scoring systems through AI and sensor fusion.

\section*{Acknowledgments}
The author would like to thank Mike Wong for his invaluable mentorship and guidance in the development of this project. His expertise in both Taekwondo and engineering was instrumental in shaping the ideas and execution of this research. Additional thanks to the University of California, Berkeley Martial Arts Program for their feedback and support.


\end{document}